\documentclass[amsmath,aps,prc,preprint,showkeys, showpacs]{revtex4}
\usepackage{graphicx}
\usepackage{epsfig}
\usepackage{amssymb}
\usepackage{dcolumn}% Align table columns on decimal point 
\usepackage{bm}% bold math

\newcommand{\be}{\begin{eqnarray}}
\newcommand{\ee}{\end{eqnarray}}
\def\bea{\begin{eqnarray}}
\def\eea{\end{eqnarray}}

\begin{document}

\title{Spin instabilities of infinite nuclear matter and effective tensor interactions}

\author{J. Navarro}
\affiliation{Instituto de F\'\i sica Corpuscular (CSIC-Universidad de Valencia), Apartado Postal 22085, E-46.071-Valencia, Spain}
\author{A. Polls}
\affiliation{Departament d'Estructura i Constituents de la Mat\`eria and Institut de Ci\`encies del Cosmos, Facultat de F\'{\i}sica, 
Universitat de Barcelona. Diagonal 647, E-08028 Barcelona, Spain}

\begin{abstract}
We study the effects of the tensor force, present in modern effective nucleon-nucleon interactions, 
in the spin instability of nuclear and neutron matter. Stability conditions of the system against  certain 
very low energy excitation modes are expressed in terms of Landau parameters. It is shown that in the spin case, the   stability conditions are equivalent to the condition derived from  the spin susceptibility, which is obtained as  the zero-frequency and long-wavelength limit of the spin response  function calculated  in the Random Phase Approximation. Zero-range forces of the Skyrme type and finite-range forces of M3Y and Gogny type are analyzed. It is shown that for the Skyrme forces considered, the tensor effects are sizeable, and  tend to increase the spin instability which appears at smaller densities than in the case that the tensor is not taken into account. On the contrary, the tensor contribution of finite range forces to the spin susceptibility is small or negligible for both isospin channels of symmetric nuclear matter as well as for neutron matter. A comparison with the spin susceptibility provided by realistic interactions is also presented.  
\end{abstract}

\maketitle

\section{Introduction}

The energy density functional theory  applied to nuclear systems has experienced quite a lot of progress in recent times. Using this method,  one can succesfully  perform systematic studies of binding energies and  one-body properties of nuclei in a very wide region of the nuclear chart \cite{bender2003}.
An often used energy density functional is the one obtained from the effective Skyrme 
interaction \cite{skyrme1956}  which has been very useful   in describing nuclear properties both for finite nuclei and also for the equation of state of nuclear and neutron matter \cite{skyrmestone}. 
Recently, and in order to improve the power  of the energy density functional derived from the Skyrme interaction, the possibility to incorporate a zero range tensor component to the effective interaction has been explored \cite{brown2006,brink2007,doba2008,colo2009}. In principle, to add tensor components to the effective nucleon-nucleon interaction seems rather natural. After all, they are a crucial ingredient of any realistic interaction,  able to predict nucleon-nucleon (NN) scattering data and deuteron properties. In fact, the tensor component of the nucleon-nucleon interaction is
 recognized as the responsible for the quadrupolar momentum of the deuteron.   
However, the effective Skyrme interactions are built to be used mainly at the mean field
 level, and in  this approach,  the tensor component gives zero contribution to the ground state binding energy of spin saturated systems, for example symmetric nuclear matter (SNM) or pure neutron matter (PNM). This fact together with the difficulties to constraint the coupling constants associated to the tensor force are  some  of the reasons for which the use of the tensor forces in the effective interactions has been delayed for a long time.
 However, one should keep in mind that the first proposed Skyrme force was already
 containing  tensor components \cite{skyrme1959} and that some early calculational
 efforts were performed almost thirty years ago \cite{stancu1977} to study the effects of the tensor component on the quasi-particle interaction at the Fermi surface of neutron matter 
\cite{pawel75}
as well as on the location of the single-particle energies of nuclei \cite{stancu1977} based on the
 Skyrme force parametrization SIII \cite{giai1975}.
Presently, the incorporation of the tensor force is becoming a standard procedure  making the effective  interaction much powerful specially in what concerns the calculations of excitation energies and response functions. 
Recently, the Lyon group  performed a systematic analysis of the effects of a zero-range tensor component, added to  a Skyrme type forces, in the response function of nuclear and neutron matter to different probes.   The study was  conducted in the framework of the Random Phase Approximation (RPA). The  responses and the corresponding dynamic structure functions
 have been further analyzed by means of energy weighted sum-rules 
 \cite{meyer2009,meyer12012,meyer22012}.
These authors have shown that the response of symmetric nuclear matter as well as of neutron matter are strongly affected by the presence of this extra term in the Skyrme interaction.

The purpose of the present paper is to perform an explicit calculation of the static spin response 
of SNM and PNM in the long wavelength limit for effective interactions containing tensor components. 
The calculation is performed in the framework of the RPA fully taking into account the exchange terms. In the next section, we recover the expressions of the  stability conditions in the spin and isospin channels  in terms of some inequalities that the Landau parameters should respect \cite{bac79}. In section III, the derivation of the spin susceptibiliy as the long wavelength limit of
 the static spin response is discussed in detail. This limit allows to express the result in terms of
 the Landau parameters. The qualitative and quantitative  analysis of the tensor effects for different effective  forces of zero and finite range is presented in section IV. Finally, a short summary of the conclusions is given in the last section. 

\section{Stability conditions and Landau parameters}
\label{SEC:Landau}

An intermediate step towards the calculation of the response of a nuclear system is to 
obtain the particle-hole (ph)  interaction. 
Expanding the interaction in Legendre polynomials $P_{\ell}(\hat{k}_1 \cdot \hat{k}_2 )$,
 the ph interaction for SNM is written as 
\bea
&& \sum_{\ell} P_{\ell}(\hat{k}_1 \cdot \hat{k}_2 ) \left\{  f_{\ell} + f'_{\ell} (\tau_1 \cdot \tau_2) +  
g_{\ell} (\sigma_1 \cdot \sigma_2)  \right. \nonumber \\
&& \left. +  g'_{\ell} (\tau_1 \cdot \tau_2) (\sigma_1 \cdot \sigma_2) +
\left[ h_{\ell} + h'_{\ell} (\tau_1 \cdot \tau_2) \right] \left. 
\frac{k_{12}^2}{k_F^2} S_{12}({\bf \hat k}_{12}) \right|_{k_i=k_F} \right\} \, ,
\label{VphLAN}
\eea
where $f_{\ell}, g_{\ell}, \dots$ are the Landau parameters, which depend on the density 
through the Fermi momentum $k_F$, and ${\bf k}_{12}={\bf k}_1-{\bf k}_2$ where ${\bf k}_1$ and
${\bf k}_2$ are the initial and final hole states affected by the ph interaction.
Notice that the particle and the hole momenta are both at the Fermi surface and therefore their 
moduli are equal to the   Fermi momentum $k_F$. The ph interaction thus depends only on the
angle $(\hat{k}_1 \cdot \hat{k}_2 )$ between the initial and final hole momenta. 
%is the relative momentum of the ph state, whose modulus ranges from 0 to 2 $k_F$.
The tensor operator in momentum space is defined as
\be
S_{12}({\bf \hat k}) = 3 (\sigma_1 \cdot {\bf \hat k})  (\sigma_2 \cdot {\bf \hat k}) -  (\sigma_1 \cdot \sigma_2) \, .
\ee  
We have followed the convention of Refs.~\cite{dab76,bac79},  for the definition of the tensor 
parameters $h_{\ell}, h'_{\ell}$.
Notice that for  PNM, the combinations $f_{\ell}+f'_{\ell} (\tau_1 \cdot \tau_2)$, $g_{\ell}+g'_{\ell} (\tau_1 \cdot \tau_2)$, and $h_{\ell}+h'_{\ell} (\tau_1 \cdot \tau_2)$
should be replaced with the coefficients $f^{(n)}_{\ell}, g^{(n)}_{\ell}, h^{(n)}_{\ell}$, respectively.
 It is customary  to deal with dimensionless parameters $F_{\ell}, G_{\ell}, \dots ,$ obtained from the previous ones by multiplying them with the density of states per energy at the Fermi  surface 
 $ N_0= n_d \frac{k_F m^*}{ 2 \hbar^2 \pi^2}$, where $n_d$ is the spin-isospin degeneracy factor, i.e. 4 for SNM and  2 for PNM. 

A first observation is the absence of the spin-orbit terms in the particle-hole interaction
at the Fermi surface. One could think that they have been omitted in the interaction. However, it is 
easily shown that spin-orbit terms, either zero- or finite-range, are proportional to the transferred momentum  and consequently do not contribute to the Landau parameters. 

As we want to concentrate on the effects of the tensor force,  we   consider only the  spin $S=1$ channel. For the sake of simplicity, in the following we will explicitly derive  our results in  the isospin channel $I=0$  of symmetric nuclear matter. However, the final expressions are also applicable to the
 $I=1$ channel or to neutron matter by replacing the parameters $G_{\ell}, H_{\ell}$ by  $G'_{\ell}, H'_{\ell}$, and $G^{(n)}_{\ell}, H^{(n)}_{\ell}$, respectively. 

The stability  of the spherical Fermi surface of nuclear matter against small deformations can be expressed in terms of the Landau parameters which should fulfill some stability criteria.
As shown in Ref.~\cite{bac79}, the inclusion of tensor components in the ph interaction
 produces a coupling between the spin-dependent parts, that is, $G_{\ell}$ terms
 are coupled to $H_{\ell}$ ones. Compact expressions are obtained considering states with
 good ph angular momentum $J$, then  the potential part of the free energy for a given
 $J$ is a $2 \times 2$ matrix with $\ell, \ell'$ values $J \pm 1$. The stability 
criterion is given by the condition that these matrices have positive eigenvalues. 
The matrices are diagonal if $\ell=\ell'$, and one gets a single stability criterion
 for each possible value of $J$.

We collect here the resulting stability criteria for the lower possible values of $\ell$ and $J$. 
The first diagonal matrices correspond to the values $\ell=1, \ell'=1$, and one gets
\bea
&& 1 + \frac{1}{3} G_1 - \frac{10}{3} H_0 + \frac{4}{3} H_1 - \frac{2}{15} H_2 > 0 
\label{IN-L1-0} \, , \\
&& 1 + \frac{1}{3} G_1 + \frac{5}{3} H_0 - \frac{2}{3} H_1 + \frac{1}{15} H_2 > 0 
\label{IN-L1-1} \, , \\
&& 1 + \frac{1}{3} G_1 - \frac{1}{3} H_0 + \frac{2}{15} H_1 - \frac{1}{75} H_2 > 0 \, ,
\label{IN-L1-2}
\eea
for $J=0^-, 1^-$ and $2^-$, respectively. The diagonal $\ell=2, \ell'=2$ case gives
\bea
&& 1 + \frac{1}{5} G_2 - \frac{7}{15} H_1 + \frac{2}{15} H_2 - \frac{3}{35} H_3  > 0 
\label{IN-L2-1}\, , \\
&& 1 + \frac{1}{5} G_2 + \frac{7}{15} H_1 - \frac{2}{5} H_2 + \frac{3}{35} H_3  > 0 
\label{IN-L2-2} \, , \\
&& 1 + \frac{1}{5} G_2 - \frac{2}{15} H_1 + \frac{4}{35} H_2 - \frac{6}{245} H_3  > 0 \, , 
\label{IN-L2-3}
\eea
for $J=1^+, 2^+$, and $3^+$, respectively. Obviously,  taking  all tensor parameters 
equal to zero, one recovers the familiar stability conditions $G_{\ell} > - (2 \ell+1)$. 

The first two non-diagonal matrices correspond to the coupling between
 ($\ell=0, J=1^+$) and ($\ell=2, J=1^+$) modes
\be
\begin{pmatrix}
1+G_0 && - \sqrt{2} ( H_0-\frac{2}{3} H_1 + \frac{1}{5} H_2) \\
 - \sqrt{2} ( H_0-\frac{2}{3} H_1 + \frac{1}{5} H_2)  && 1+\frac{1}{5} G_2 
 - \frac{7}{15} H_1 + \frac{2}{5} H_2 - \frac{3}{35} H_3
 \end{pmatrix} \, ,
\label{MAT-J1}
 \ee
and to the coupling between ($\ell=1, J=2^-$) and ($\ell=3, J=2^-$):
\be
\begin{pmatrix}
1+\frac{1}{3} G_1-\frac{1}{3} H_0 + \frac{2}{15} H_1 - \frac{1}{75} H_2 &&
- \sqrt{6} (\frac{1}{5} H_1 - \frac{6}{25} H_2 + \frac{3}{35} H_3) \\
- \sqrt{6} (\frac{1}{5} H_1 - \frac{6}{25} H_2 + \frac{3}{35} H_3) &&
1 + \frac{1}{7} G_3 - \frac{36}{175} H_2 + \frac{8}{35} H_3 - \frac{4}{63} H_4
\end{pmatrix} \, .
\label{MAT-J2}
\ee
Later on, the two sets of inequalities (\ref{IN-L1-0}--\ref{IN-L1-2}), (\ref{IN-L2-1}--\ref{IN-L2-3}), and the two stability criteria associated to matrices (\ref{MAT-J1}) and (\ref{MAT-J2}) will be referred to as 
$\ell=1$, $\ell=2$, $J=1$ and $J=2$ sets, respectively. 
In Sect.~\ref{SEC:RES}, we shall discuss the behavior of these four sets of inequalities
 as a function of the density for several effective interactions. In the next section we are going to show that the stability criteria (\ref{MAT-J1}) is in fact related to the spin susceptibility.

\section{The spin susceptibility}
\label{SEC:SUSC}

In general, the susceptibility of a system is given by the  linear response function to the appropriate 
external probe.  We  use the notation $\chi^{(\alpha)}_{RPA}(\bf{q},\omega)$ for the response function calculated in the RPA, and $\chi^{(\alpha)}_{RPA}(0)$ for the corresponding 
zero-frequency and  long-wavelength limit defining the susceptibility. The same quantities in the mean-field or Hartree-Fock approximation will be indicated with the subindex (HF). 
The symbol $(\alpha)$ refers to the quantum numbers of the ph spin-isospin channel $(S,M;I,Q)$, with $M$ and $Q$ being the projections on the selected quantization axis of the spin and isospin operators, respectively. The spin response is linked to the  $S=1$ channel, without dynamical terms mixing isospin, and no charge exchange. Therefore only the quantum number $M$ is relevant in the following. 

We employ the method and notations presented in Refs. \cite{gar92,mar06,meyer2009} to obtain the RPA response function. The starting point is the Bethe-Salpeter equation for the Green's function, or retarded ph propagator 
\bea
G_{RPA}^{(M)}({\bf q},\omega,{\bf k}_1) &=& 
G_{HF}({\bf q},\omega,{\bf k}_1) \nonumber \\
&+& G_{HF}({\bf q},\omega,{\bf k}_1) 
\sum_{M'} \int \frac{d^3 k_2}{(2 \pi)^3} V_{ph}^{(M,M')}({\bf q}, {\bf k}_1, {\bf k}_2)  
G_{RPA}^{(M')}({\bf q},\omega,{\bf k}_2) \, ,
\eea
where 
\be
G_{HF}({\bf q},\omega,{\bf k}_1) = 
\frac{\theta(k_F-k_1) - \theta(k_F-|{\bf k}_1 + {\bf q}|)}{\omega+\varepsilon(k_1) -
\varepsilon(|{\bf k}_1 + {\bf q}|) + i \eta} \, ,
\ee
is the Hartree-Fock ph propagator, which is independent on the specific $M$-channel, 
 and $V_{ph}^{(M,M')}$ refers to the matrix elements of the ph interaction. 
In this notation, $\omega$ is the transferred energy, ${\bf k}_1$ and ${\bf k}_2$ are the initial and final hole momenta in the ph interaction, respectively, and  ${\bf q}$ is the transferred momentum.

The response function in the homogeneous medium is related to the ph Green's function by
\be
\chi_{RPA}^{(M)}({\bf q},\omega) = n_d \int \frac{d^3 k_1}{(2 \pi)^3}
G_{RPA}^{(M)}({\bf q},\omega,{\bf k}_1) \, .
\label{chiRPA}
\ee
The HF response function $\chi_{HF}$, or Lindhard function, is obtained when the free propagator $G_{HF}$ is used in Eq.~(\ref{chiRPA}). 
In the following we will often deal with averages  similar to the one appearing  in Eq.~(\ref{chiRPA}),
that will be denoted with  the  notation $\langle G_{RPA}^{(M)}({\bf k}_1) \rangle$; the integrated momentum ${\bf k}_1$ will  be dropped when there is no confusion.

We now specify a  ph interaction described in terms of Landau parameters, as the one defined in Eq.~(\ref{VphLAN}), and limited to the $S=1, I=0$ space. The matrix elements of these ph interaction  in the spin-isospin space are written as
\bea
V_{ph}^{(M,M')}(1,2) &=&  \delta(M,M') n_d \sum_{\ell}  {g}_{\ell} P_{\ell}(\hat{k}_1 \cdot
 \hat{k}_2)  + n_d \sum_{\ell}  {h}_{\ell} P_{\ell}(\hat{k}_1 \cdot \hat{k}_2)  S_T^{(M,M')}(
{\bf \hat k}_1,{\bf \hat k}_2) \, ,
\label{Vph}
\eea
where  
\be
S_T^{(M,M')}({\bf \hat k}_1,{\bf \hat k}_2) = 3 (-)^M \left(k_{12} \right)^{(1)}_{-M} \left( k_{12} \right)^{(1)}_{M'}
-  \delta(M,M')  2 [ 1 -  (\hat{k}_1 \cdot \hat{k}_2) ] \, .
\ee
Following the notation of Ref.~\cite{meyer2009} we have defined
\be
\left(k_{12} \right)^{(1)}_{M} = \sqrt{\frac{4 \pi}{3}}
 \left( Y_{1,M}(\hat{k}_1) - Y_{1,M}(\hat{k}_2) \right) \, .
 \ee

For practical purposes it is convenient to expand  the Legendre polynomials and collect the powers of $\hat{k}_1 \cdot \hat{k}_2$. The matrix elements of the ph interaction are thus written as
\bea
V_{ph}^{(M,M')}(1,2) &=& \delta(M,M') \sum_n A_n (\hat{k}_1 \cdot \hat{k}_2)^n 
+ \sum_n B_n  (\hat{k}_1 \cdot \hat{k}_2)^n  
(-)^M \left(k_{12} \right)^{(1)}_{-M} \left( k_{12} \right)^{(1)}_{M'} \, ,
\label{Vph-AB}
\eea
where $A_n$ and $B_n$ are combinations of Landau parameters and coefficients of Legendre polynomials. 
The Bethe-Salpeter equation can then be written as:
\bea
G_{RPA}^{(M)}({\bf q},\omega,{\bf k}_1)  &=& G_{HF}({\bf q},\omega,{\bf k}_1) + G_{HF}({\bf q},\omega,{\bf k}_1)  \sum_n A_n 
\langle (\hat{k}_1 \cdot \hat{k}_2)^n G_{RPA}^{(M)}({\bf q},\omega,{\bf k}_2) \rangle
\nonumber \\
 &+&
G_{HF}({\bf q},\omega,{\bf k}_1) \sum_n B_n  \sum_{M'} \langle (\hat{k}_1 \cdot \hat{k}_2)^n 
(-)^M \left(k_{12} \right)^{(1)}_{-M} \left( k_{12} \right)^{(1)}_{M'}
  G_{RPA}^{(M')}({\bf k}_2) \rangle \, .
\label{BS}
  \eea
Integrating over the momentum ${\bf k}_1$ one gets the equation for  the response function  
\bea
\chi_{RPA}^{(M)}({\bf q},\omega)  &=& \chi_{HF}({\bf q},\omega) + \sum_n A_n K_1(n,M) \nonumber \\
&+&  \sum_n B_n  \left\{ K_2(n,M) - K_3(n,M) - K_4(n,M) + K_5(n,M) \right\} \, ,
\label{lachi}
\eea
where we have defined the functions $K_i(n,M)$ which involve an average  over the hole momenta ${\bf k}_1$ and ${\bf k}_2$
\be
K_i(n,M) = \sum_{M'} \langle \, G_{HF}({\bf q},\omega,{\bf k}_1) \, 
  (\hat{k}_1 \cdot \hat{k}_2)^n \, Q_i \, G_{RPA}^{(M')}({\bf q},\omega,{\bf k}_2)  \rangle \, ,
\label{K1-5}
\ee
including one of the five quantities:
\bea
Q_i &=& \left\{\delta(M,M'), \frac{4 \pi}{3} Y_{1,M}^*(\hat{k}_1) Y_{1,M'}(\hat{k}_1), \frac{4 \pi}{3} Y_{1,M}^*(\hat{k}_1) Y_{1,M'}(\hat{k}_2), \right. \nonumber \\
&& \left. \hspace{1cm}  \frac{4 \pi}{3} Y_{1,M}^*(\hat{k}_2) Y_{1,M'}(\hat{k}_1), \frac{4 \pi}{3} Y_{1,M}^*(\hat{k}_2) Y_{1,M'}(\hat{k}_2) \right\} \, .
\eea
$Q_1$ is related to the term multiplying the coefficient $A_n$ in Eq.~(\ref{BS}), and the  remaining $Q_i$ to the four contributions involving the coefficients $B_n$. Writing now 
\be
({\hat k}_1 \cdot \hat{k}_2)^n = \left( \frac{4 \pi}{3} \right)^n \sum_{\mu_1 \dots \mu_n}
 Y^*_{1,\mu_1}(\hat{k}_1) \dots Y^*_{1,\mu_n}(\hat{k}_1) 
 Y_{1,\mu_1}(\hat{k}_2) \dots Y_{1,\mu_n}(\hat{k}_2) \, ,
\ee
one can see that the averages  entering the definitions (\ref{K1-5}) of the functions $K_i(n,M)$ are in fact the product of two integrals: one over the momentum ${\bf k}_1$ involving the propagator $G_{HF}$, and the other over the momentum ${\bf k}_2$ involving the propagator $G_{RPA}$. Each of these integrals contain also products of spherical harmonics of order 1 coming from $Q_i$ and from $({\hat k}_1 \cdot \hat{k}_2)^n$.
 
The response function is thus coupled to other integrals of the RPA Green's function weighted with products of spherical harmonics $Y_{1,m}(\hat{k})$. For instance, one of such integrals correspond to the function $K_5(0,M)$, so that the response function is coupled to 
\bea
S^{(M)}({\bf q},\omega) = n_d \frac{4 \pi}{3} \sum_{M'}  \langle Y^*_{1,M}(\hat{k}) Y_{1,M'}(\hat{k}) 
G^{(M'')}_{RPA}({\bf q}, \omega, {\bf k}) \rangle  \, .
\eea
To write the equation satisfied by this function one should follow the next steps: go back to the 
Bethe-Salpeter Eq.~(\ref{BS}), multiply both sides with the weighting factor entering the definition of
 $S^{(M)}({\bf q},\omega)$, take the sum over $M'$ and finally integrate over the hole momentum, to get 
\bea
S^{(M)}({\bf q},\omega)  &=& \frac{1}{3} \chi_{HF}({\bf q},\omega) + \sum_n A_n K_2(n,M) \nonumber \\
&+&  \sum_n B_n  \left\{ K_2(n,M) - K_3(n,M) - K_2(n+1,M) + K_5(n+1,M) \right\} \, .
\label{laese}
\eea
To obtain the full response function one should proceed in a similar way and write more coupled equations until a closed algebraic system is obtained. In principle, one should consider an infinite number of equations, and the solution of the system seems not easy to obtain. Only restricting the number of Landau parameters, one can expect to get suitable expressions. 

However, as we are only interested in getting the spin susceptibility, we have to calculate these integrals in the zero-frequency and long-wavelength limit, for which the left hand side of Eqs.~(\ref{lachi}) and (\ref{laese}) will be written with the short notation $\chi^{(M)}_{RPA}(0)$ and $S^{(M)}(0)$, with a single null argument. 
As shown in the Appendix, it turns out that in that limit the quantities $K_i(n,M)$ are linear combinations of only $\chi^{(M)}_{RPA}(0)$ and $S^{(M)}(0)$, and consequently 
Eqs.~(\ref{lachi}) and (\ref{laese}) form a closed algebraic system.

Finally, in terms of dimensionless Landau parameters the coupled system writes as:
\be
\left( 1 + G_0-H_0+\frac{2}{3} H_1-\frac{1}{5} H_2 \right) \chi_{RPA}^{(M)}(0) +
\left( 3 H_0-2H_1+ \frac{3}{5} H_2 \right) S^{(M)}(0) = \chi_{HF}(0)  
\ee
\bea
\left( \frac{1}{3} G_0-\frac{1}{15} G_2 + \frac{1}{3} H_0-\frac{1}{15} H_1-\frac{1}{15} H_2 + \frac{1}{35} H_3 \right)
\chi_{RPA}^{(M)}(0)  && \nonumber \\
+ \left( 1+ \frac{1}{5} G_2+H_0-\frac{17}{15} H_1+\frac{3}{5} H_2-\frac{3}{35} H_3 \right) S^{(M)}(0) &=&
\frac{1}{3} \chi_{HF}(0) \,.
\eea

Solving this system, the spin susceptibility can be expressed in terms of the Landau parameters:
\bea
\frac{\chi_{HF}(0)}{\chi_{RPA}^{(M)}(0)} &=& 1 + G_0 - 
\frac{2 \, (H_0 - \frac{2}{3} H_1 + \frac{1}{5} H_2)^2 }{1+\frac{1}{5} G_2-\frac{7}{15} H_1 + \frac{2}{5} H_2 - \frac{3}{35} H_3} \, .
\label{suscept}
\eea
Notice that the spin susceptibility is independent of the spin projection $M$, {\it i.e.} the spin susceptibility is identical for the longitudinal ($M=0$) and the transverse ($M=1$) channels. Obviously, it reduces to the familiar spin susceptibility of a Fermi liquid when no tensor terms ($H_{\ell}=0$) are considered. As mentioned above, this expression for the susceptibility is also valid both for the isovector channel $I=1$ of SNM and for PNM, just replacing the set of parameters $(G_{\ell}, H_{\ell})$ with  $(G'_{\ell}, H'_{\ell})$ and $(G^{(n)}_{\ell}, H^{(n)}_{\ell})$, respectively.

In Ref. \cite{fuj87} Fujita and Quader extended the original Landau theory of Fermi liquids to a ph interaction invariant under a combined rotation in spin and  orbital spaces to study electron systems and heavy fermions. These authors considered a general spin interaction, including the ordinary tensor one. Solving the kinetic Boltzmann equation they derived the spin susceptibility in terms of Landau parameters. It turns out that keeping only parameters $H_{\ell}$, their expression 
coincides with Eq. (\ref{suscept}). Along similar lines Olsson {\it et al.} ~\cite{ols04} 
calculated also the spin susceptibility of pure neutron matter starting from the kinetic equation.
 Besides the usual tensor contribution, these authors also included two new terms, namely the
 center of mass tensor and the crossed vector tensor. In addition, they also used a different 
definition of the tensor Landau parameters. Although they do not provide a final experssion for 
the spin susceptibility in terms of the Landau parameters we have checked that taking into account 
the different definitions employed for the Landau parameters and not considering the terms 
of the center of mass tensor and the crossed vector tensor one gets an expression in agreement 
with Eq. (\ref{suscept}).  

Let us turn now to the stability criteria provided by the $J=1$ set, which states
 that the eigenvalues of the matrix (\ref{MAT-J1}) should be positive. For the sake
 of having a simple notation, let us denote by $M_{11}$, $M_{12}=M_{21}$ and $M_{22}$ its matrix elements. Then, the  two eigenvalues of matrix (\ref{MAT-J1}) can be written as
\be
\lambda_{\pm} = \frac{1}{2} \left( M_{11} + M_{22} \right) \pm \frac{1}{2} \sqrt{ \left( M_{11}-M_{22} \right)^2 + 4 M_{12}^2} \, .
\label{eigen}
\ee
Obviously $\lambda_{+} \ge \lambda_{-}$, so in practice the critical density associated to the $J=1$ set is defined by the condition $\lambda_-=0$. The spin susceptibility provides an additional stability criterion, namely for a stable system it should be positive for all densities, otherwise it undergoes a spontaneous ferromagnetic transition. The related critical density corresponds to the density value for which $1/ \chi_{RPA}(0) = 0$. Note that the inverse spin susceptibility (\ref{suscept}) can be written in terms of the matrix elements $M_{ij}$ of (\ref{MAT-J1}) as 
\be
\frac{\chi_{HF}(0)}{\chi_{RPA}^{(M)}(0)} =M_{11} - M_{12}^2/M_{22} \, . 
\label{eigensus}
\ee
Using Eqs. (\ref{eigen}) and (\ref{eigensus}), it is straightforward to  shown that the 
equalities $\lambda_{-} \,  = 0$ and $1/\chi_{RPA}^{(M)}(0)=0$ are equivalent. Therefore, both conditions define the same critical density, which is the relevant physical output related 
to both stability criteria.

 \section{Results}
\label{SEC:RES}

\subsection{Skyrme interactions}

The zero-range tensor force employed in the current effective interactions is the one originally proposed by Skyrme \cite{skyrme1959}, and has the following structure:
\begin{eqnarray}
V^T &=& \frac{t_e}{2} \left\{ \left[ 3 (\sigma_1 \cdot {\bf k'} ) (\sigma_2 \cdot {\bf k'}) -  (\sigma_1 \cdot \sigma_2)
{\bf k'}^2 \right] \delta ( {\bf r}_1 - {\bf r}_2) + h.c \right\} 
  \\
 &+& \frac {t_o}{2} \left\{ \left [3 (\sigma_1 \cdot {\bf k'}) \delta({\bf r}_1 - {\bf r}_2 ) (\sigma_2 \cdot {\bf k}) -  \left[
(\sigma_1 \cdot \sigma_2) {\bf k}' \cdot \delta({\bf r}_1 - {\bf r}_2) {\bf k} \right] \right ]  + h.c. \right\} \, ,
\end{eqnarray}
where the operator  ${\bf k} = ( \overrightarrow{\nabla}_1 - \overrightarrow{\nabla}_2)/2i$ acts on the ket  while its adjoint  $ {\bf k}' =  - ( \overleftarrow{\nabla}_1 - \overleftarrow{\nabla}_2)/2i$ acts on the bra. The coupling constants $t_e$ and $t_o$ measure the strength of the interaction in the triplet-even and triplet-odd channels, respectively. It only contributes to the $\ell =0$ Landau parameters, which have a simple expression in terms of the coupling constants of the interaction:
\be 
H_0 = N_0 \frac {k_F^2}{8}  [ t_e + 3 t_o] \, ,
\label{eq:ho}
\ee
\be
H_0' = N_0 \frac {k_F^2}{8}  [ -t_e + t_o  ] \, ,
\label{eq:hop}
\ee
for symmetric nuclear matter and 
\be
H_0^{(n)} = N_0 \frac {k_F^2}{2} t_o \, ,
\label{eq:hon}
\ee
for neutron matter.

Therefore, for the analysis of the spin-instabilities the only non-vanishing Landau parameters are 
$(G_0, G_1, H_0)$, so that the $\ell=2$ set does not affect the stabilities.
In fact, only the $\ell=1$ and $J=1$ sets of inequalities (Eqs. \ref{IN-L1-0}-\ref{IN-L1-2})  are relevant. The former writes 
\bea
 J=0^-   &:&   1 + \frac {1}{3} G_1 - \frac {10}{3} H_0   > 0  \, ,\\
J= 1^-   &:&   1 + \frac {1}{3}  G_1 + \frac {5}{3} H_0   > 0 \, ,\\
J= 2^-  &:&  1+ \frac {1}{3} G_1 - \frac {1}{3} H_0    > 0  \, .
\eea
It is worth noting that the lower bound of these inequalities is provided either by the $J=0^-$ one if $H_0 \ge 0$ or by the $J=1^-$ one if $H_0 \le 0$. Thus the $J=2^-$ inequality plays no role in the discussion of instabilities. Concerning the $J=2$ set, by simple inspection of the matrix (\ref{MAT-J2}) one can see that its eigenvalues are $1 + G_1/3 - H_0/3$ and $1$. The condition on the former is identical to the previous $J=2^-$ inequality.   

The stability conditions associated to the matrix (\ref{MAT-J1}) are identical to the condition on the spin susceptibility (\ref{suscept}), which is written as
\bea
\frac{\chi_{HF}(0)}{\chi_{RPA}^{(M)}(0)} &=& 1 + G_0 
-2 H_0^2 \, .
\label{suscSKY}
\eea
This result is particularly relevant for pure neutron matter. Indeed, currently used Skyrme
 interactions predict that even in the absence of a magnetic field a spontaneous magnetization arises in pure neutron matter at a critical density which can be as low as $\simeq 1.1\rho_0$~\cite{vid84,kut89,isa04,vida05}, depending on the specific parameterization. 
Eq. (\ref{suscSKY}) shows that the inclusion of the tensor interaction results always in a  smaller value for the critical density  than the one without tensor terms, and consequently the spontaneous magnetization would appear at densities closer to $\rho_0$. In the next subsection we shall show that neither finite-range effective interactions \cite{lopez2006}  nor realistic calculations \cite{ramos2002} predict such a spontaneous magnetization in neutron matter, which therefore should be considered as an artifact of the  Skyrme interaction. 

In Ref.~\cite{colo2010}, the $\ell=1$ and $J=1$ sets of stability conditions have been analyzed for symmetric nuclear matter in the spin-isospin channels (1,0) and (1,1). Critical densities were determined for a variety of Skyrme interactions with tensor contributions whose strength coupling constants have been fixed in two different ways. In some cases, the tensor force is simply added to an existing Skyrme parameterization and the tensor parameters have been 
fitted to reproduce some specific property. In other cases, tensor parameters have been
 incorporated to the effective force and the fitting procedure is started over,  as in the case of the T$IJ$ set~\cite{les07}. Interestingly in 28 out from the 41 analyzed parameterizations in Ref. ~\cite{colo2010}, the smallest  critical density is related to the violation of the $J=1$ stability criteria, that is by the condition of Eq. (\ref{suscSKY}), for the isospin $I=1$ channel in symmetric nuclear matter.

\begin{figure}[ht]
\centering\includegraphics[width=\linewidth]{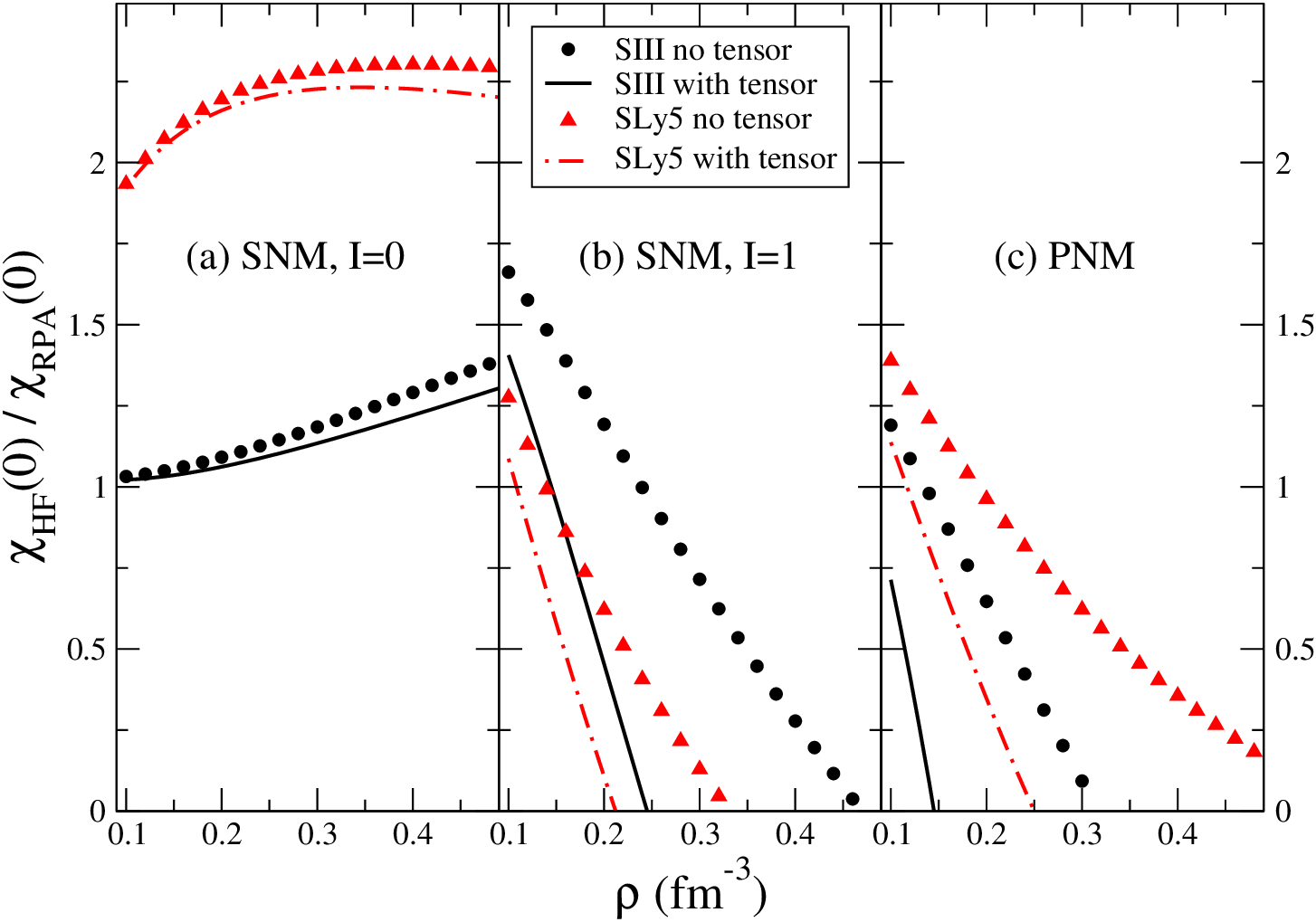}
\caption{(Color online) Inverse of the spin susceptibility respect to the inverse Hartree-Fock spin susceptibility as a function of density for the Skyrme interactions SIII (the solid line includes the tensor effects and  the full circles are  without) and SLy5 (the dot-dashed line include the tensor effects and the triangles are without). Panels (a) and (b) correspond to symmetric nuclear matter in the singlet and triplet isospin channels respectively. Panel (c) shows the results for neutron matter.}
\label{FIG1}
\end{figure}

\begin{figure}[ht]
\centering\includegraphics[width=\linewidth]{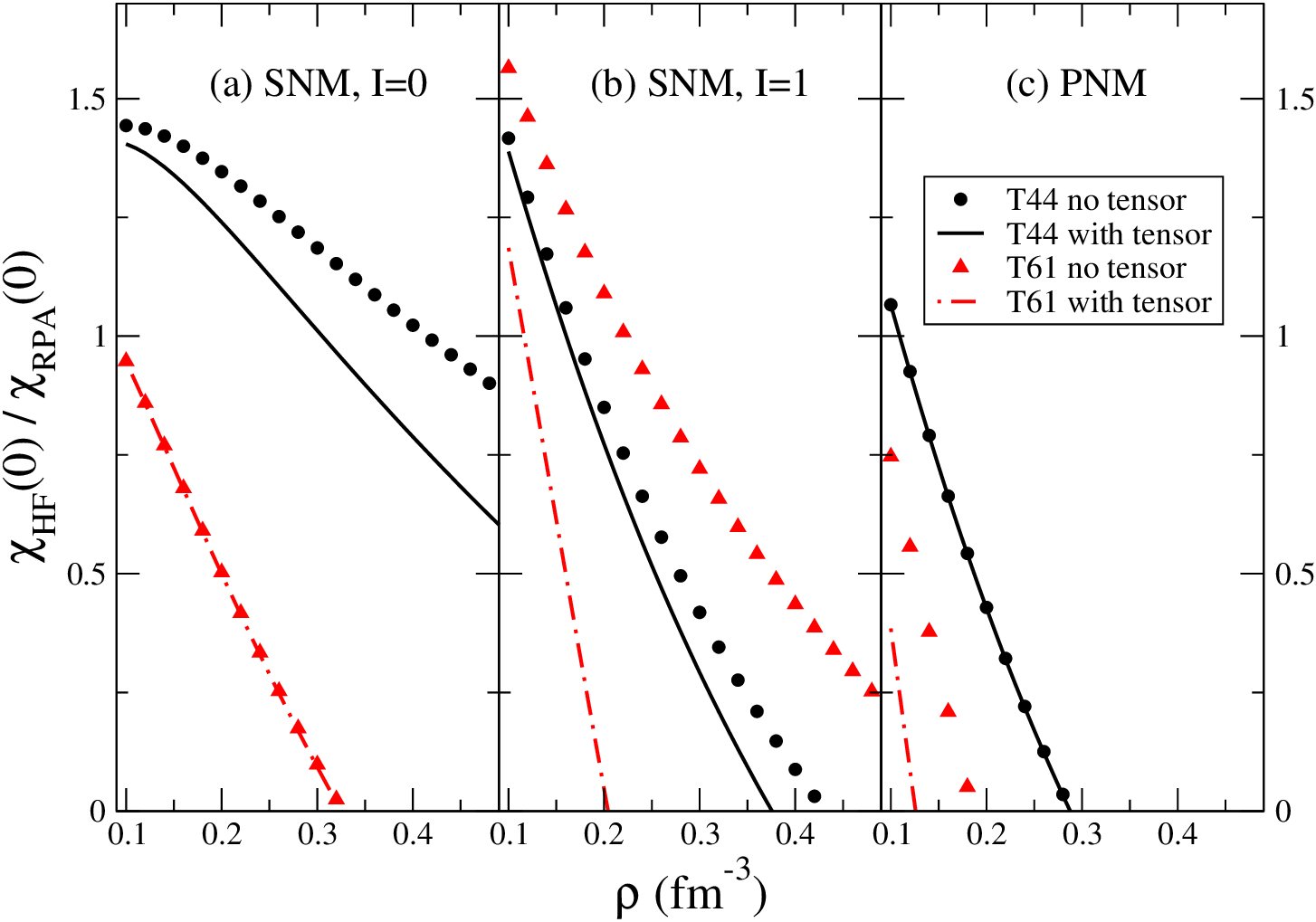}
\caption{(Color online) Inverse of the spin susceptibility respect to the inverse Hartree-Fock spin susceptibility as a function of density for the Skyrme interactions T44 (the solid line includes the tensor effects and full circles are without)  and T61 (the dot-dashed line include the tensor effects and the triangles are without). Panels (a) and (b) correspond to symmetric nuclear matter in the singlet and triplet isospin channels respectively. Panel (c) shows the results for neutron matter.}
\label{FIG2}
\end{figure}

A few illustrative examples are presented in Figs.~\ref{FIG1} and \ref{FIG2}, where  the inverse
 of the spin susceptibility is plotted as a function of the density. The isoscalar ($I=0$)
 and isovector ($I=1$) channels for SNM are shown in  panels (a) and (b), respectively.
  While, the PNM inverse susceptibility is  displayed on the  panel (c). Fig.~\ref{FIG1} shows the results for the Skyrme interaction SIII~\cite{giai1975} as well as for  SLy5~\cite{cha98}, adding in both cases the tensor terms of Ref.~\cite{col07} without modification of the original SIII and SLy5 parameters. 
The effects of the tensor force in the spin  susceptibility for the isoscalar channel are rather small in the 
range of densities considered. No signal of instability in  this channel is  detected.  However, the situation is different in the isovector channel,  where the tensor effects are rather large and the ratio $\chi_{HF}(0) /\chi_{RPA}(0)$ crosses  zero at a lower density than the case in which  the tensor is not considered. 
The  presence of the tensor component enhances the problem of spin  instability with the 
Skyrme forces. The conclusions are similar in the case of PNM, which  becomes ferromagnetic at a much lower density. 

Results for the  T$IJ$ Skyrme  forces corresponding to the parameterizations T44 and T61 are shown in Fig.~\ref{FIG2}. In these interactions the tensor is  taken into account in the global fitting procedure of the force parameters~\cite{les07}. The  tensor effects on the force T44 are rather small and for the case of SNM are similar for both isospin channels. 
The isospin channel $I=1$ shows an instability that,  as predicted,  appears at lower densities
 when the tensor is taken into account.  For PNM the effect of the tensor is very small and not
 appreciable in the figure. T44 produces a ferromagnetic transition in PNM around
 $\rho \sim 0.3 $fm$^{-3}$. In the case of T61, the tensor effects in the SNM $I=0$ channel
 are negligible. However, in the $I=1$ channel are very large and the spin instability 
appears at much lower densities. The same is true in PNM.  In conclusion, the incorporation
 of the tensor force to the Skyrme forces enhances the problems related to the
 appearance of the spin instability which appear always  at lower densities than in 
the case without tensor component. The quantitative effects depend on the specific
 force considered through the combinations of the tensor parameters of the force entering 
Eqs.~(\ref{eq:ho}-\ref{eq:hon}).
For instance, in the case of PNM, 
the small tensor effects for 
the T44 interaction are a direct consequence of  the small value of $H_0^{(n)}$ (see Table I, for $k_F= 1.7$ fm$^{-1}$)
which in turn is a reflection of the very small value of  $t_o$. Looking at the values
of $H_0^{(n)}$ for the other forces permits a quantitative understanding  of the 
observed behavior of the
different Skyrme interactions.
Similar arguments, taking the proper combinations of $t_e$ and $t_o$ clarify the
size of the tensor contributions to the spin susceptibility of nuclear matter in the
different isospin channels.

\subsection{Finite-range interactions}

We analyze  two types of finite range effective interactions, both defined in $r-$space, that use either Yukawa or Gaussian functions for the radial dependence of the interaction. Nakada~\cite{nakada2003} has constructed an effective density dependent finite range force including tensor and spin-orbit components. In this case, the tensor force is defined in $r-$space as
 $S_{12} = 3 (\vec \sigma_1 \hat r_{12}) ( \vec \sigma_2 \hat r_{12}) - \vec \sigma_1 \vec \sigma_2 $. 
 This interaction is a modification of the so called three-range Yukawa (M3Y) interaction that was originally derived from a bare  NN interaction by fitting Yukawa functions to a G-matrix \cite{bertsch1977}. In the present paper we use the parameterization M3Y-P2, which has been shown to be appropriate both for calculations in uniform systems (nuclear  and neutron matter) and finite nuclei \cite{nakada2003}. 

We also consider the traditional Gogny forces D1S~\cite{ber91} and D1M~\cite{gor09} that have been recently supplemented with a tensor force by Anguiano et al.~\cite{lallena2011}. These authors have added the tensor isospin term of the so called Argonne Av8'~\cite{wiringa2002}. This potential is built by removing operatorial components of the full Argonne Av18 potential~\cite{wiringa95}, while the remaining terms are readjusted to preserve as many lowest-order phase shifts and deuteron properties as possible. The density dependence of the energy per particle produced by this family of potentials for nuclear and neutron matter using different many-body techniques has been recently discussed~\cite{baldo2012}.  In addition, the radial part of this tensor component was  modified by multiplying the radial function of the isospin tensor component of Av8' by a factor $1-\exp (-b r_{12}^2)$ which takes care of the short-range NN correlations~\cite{lallena2011}. 
The tensor component of Argonne Av8'  has a strong short-range structure that prevents from using the potential in mean field calculations. The multiplicative factor has been shown to be an efficient way to deal with short-range NN correlations and the resulting effective interaction  has been successfully used in HF and RPA calculations of finite nuclei. The parameter $b$ has been adjusted to fit the lowest $0^-$ states of some selected nuclei in a HF plus RPA description~\cite{lallena2011}. 

\begin{figure}[ht]
\centering\includegraphics[width=\linewidth]{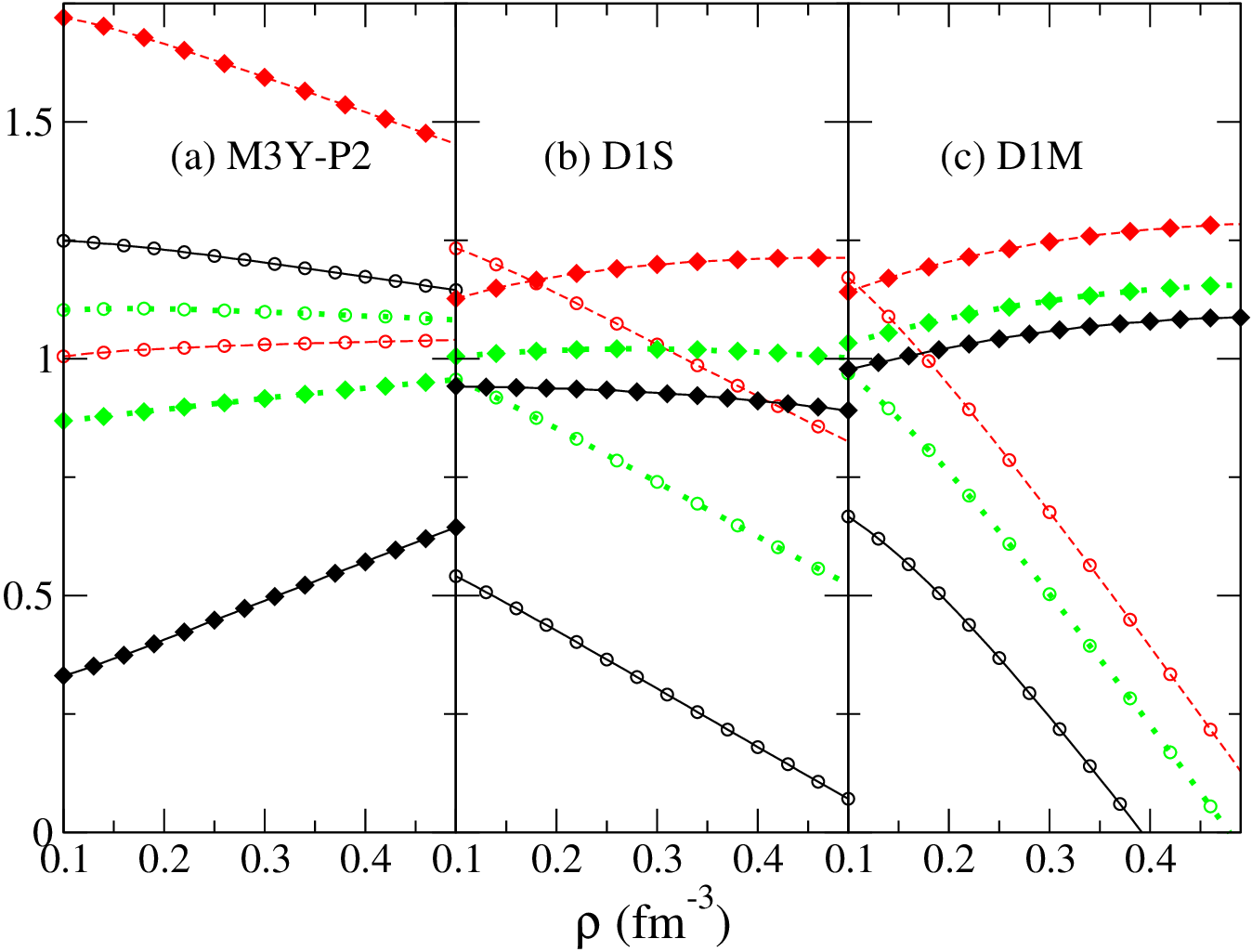}
\caption{(Color online) Instability inequalities of Eqs.(\ref{IN-L1-0}--\ref{IN-L2-3})  for the singlet isospin channel as a function of density for symmetric nuclear matter obtained for the finite range interactions: (a) M3Y-P2, (b) Gogny-D1S and (c) Gogny-D1M. Solid, dashed and dotted lines all with empty circles correspond to $L=1,  J=0^-,1^-$ and $2^-$ respectively. Solid, dashed and dotted lines all with full diamonds correspond to $L=2, J=1^+, 2^+$ and $3^+$. The effects of the tensor forces are included in all cases.}
\label{FIG3}
\end{figure}

In Fig.~\ref{FIG3}  we show the density dependence of the diagonal stability conditions in the isoscalar channel, corresponding to  $l=1$, for $J=0^-, 1^-, 2^-$ (Eqs.~\ref{IN-L1-0}--\ref{IN-L1-2}) and to $l=2$ for $J= 1^+, 2^+, 3^+$ (Eqs.~\ref{IN-L2-1}--\ref{IN-L2-3}) for symmetric nuclear matter. 
 The (a) panel shows the results for M3Y-P2 force and the (b) and (c)  panels correspond to D1S and D1M Gogny forces, respectively,  supplemented with a  tensor force as explained above. The stability conditions $l=2$ are well respected for all interactions in the density range considered in the figure. However, the $l=1$ stability condition has a negative slope as a function of density,
in particular  for the D1M Gogny force and the condition  $l=1,J=0^-$ (empty circles joint by 
 solid line) signals an instability with a critical density close to $0.4$ fm$^{-3}$.
It is worth to mention, that in the case of the isovector channel, the stability conditions are all well
 respected in the range of densities explored. The same is true for neutron matter,
whose stability conditions show a smooth behavior as a function of density, being the minimum
value about 0.75 which in the scale of the magnitudes we are analyzing can be safely
 considered far from zero. Therefore we do not show these conditions  
explicitly.

\begin{figure}[ht]
\centering\includegraphics[width=\linewidth]{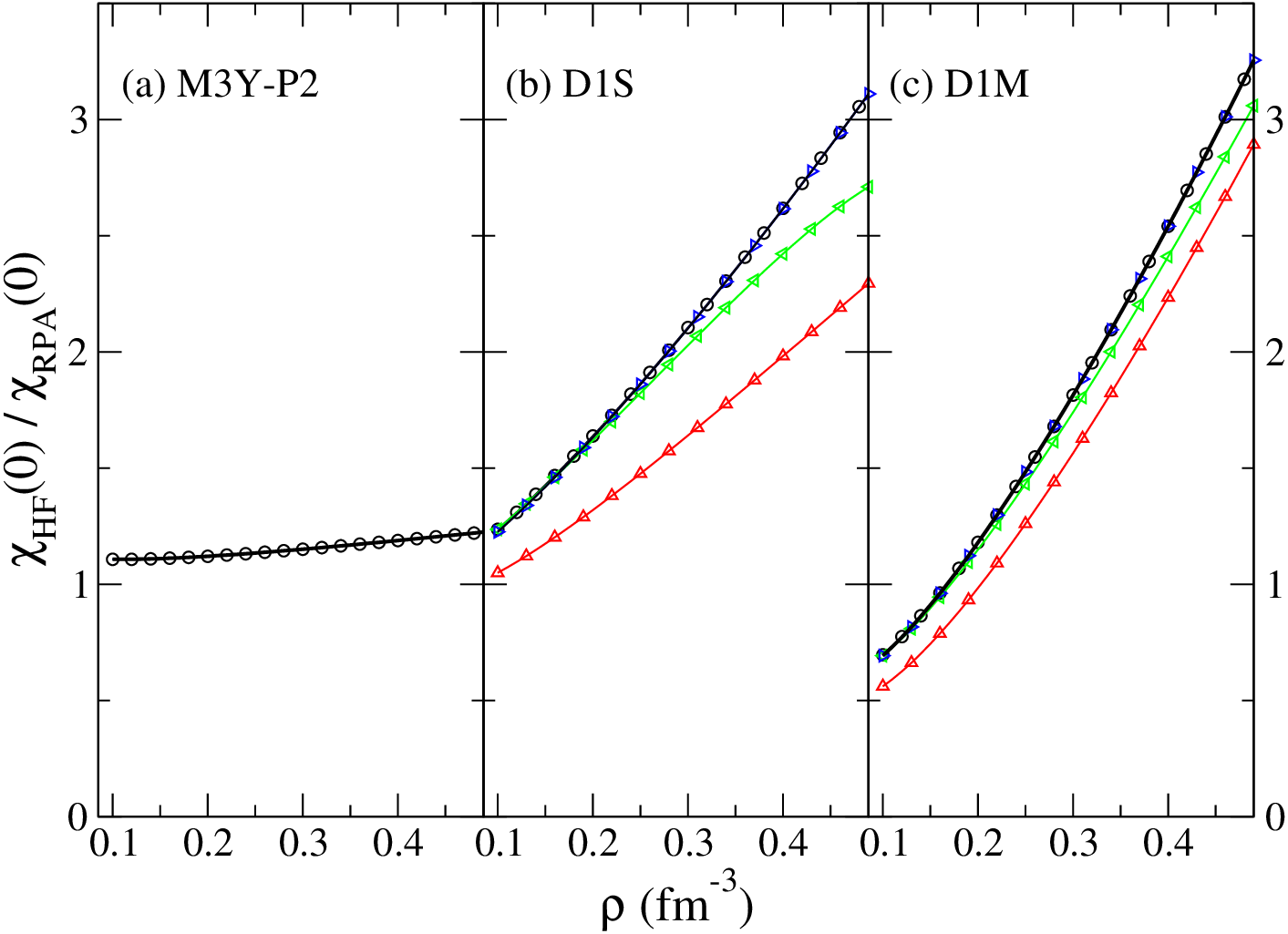}
\caption{Inverse of the spin susceptibility respect to the inverse Hartree-Fock spin susceptibility in the singlet isospin channel for symmetric nuclear matter as a function of density for the finite range effective
 forces including the tensor component: (a) M3Y-P2, (b) Gogny D1S and (c) Gogny D1M.
Empty circles stand for the results taking into account only $G_0$ in the
Eq.\ref{suscept} , with all $H$'s equal to zero. Then the different curves incorporate step by step the effects of the  $H$'s. Curve with up triangles ($G_0, H_0$), triangle oriented to the left ($G_0, H_0,H_1$), triangles oriented to the right ($G_0,H_0,H_1,H_2$) and the solid line stands for
the full $RPA$ calculation.}
\label{FIG4}
\end{figure}

In Fig.~\ref{FIG4}   we report the inverse of the nuclear matter spin susceptibility in the isoscalar channel for the same three effective interactions. To see the effect of different order Landau parameters we report the results obtained with Eq. (\ref{suscept}) in which we turn on successively the different Landau parameters. Empty circles indicate the results obtained with only $G_0$. The  curves incorporate the effect of the different $H_{\ell}$'s when they are turned on (see the figure caption). No signals of instability are observed in any of these  effective forces. In the M3Y-P2 case, the results are not affected by the Landau parameters associated to the tensor force and  basically coincide with the ones obtained by just considering the central Landau parameter $G_0$.  For the Gogny forces, the final results with all the Landau parameters involved are very close to the results obtained by using only  $G_0$. The largest deviations from the full results are obtained 
by using $G_0,H_0$. Therefore one should use the  Landau parameters up to $\ell=3$ to obtain a meaningful converged result.   

\begin{figure}[ht]
\centering\includegraphics[width=\linewidth]{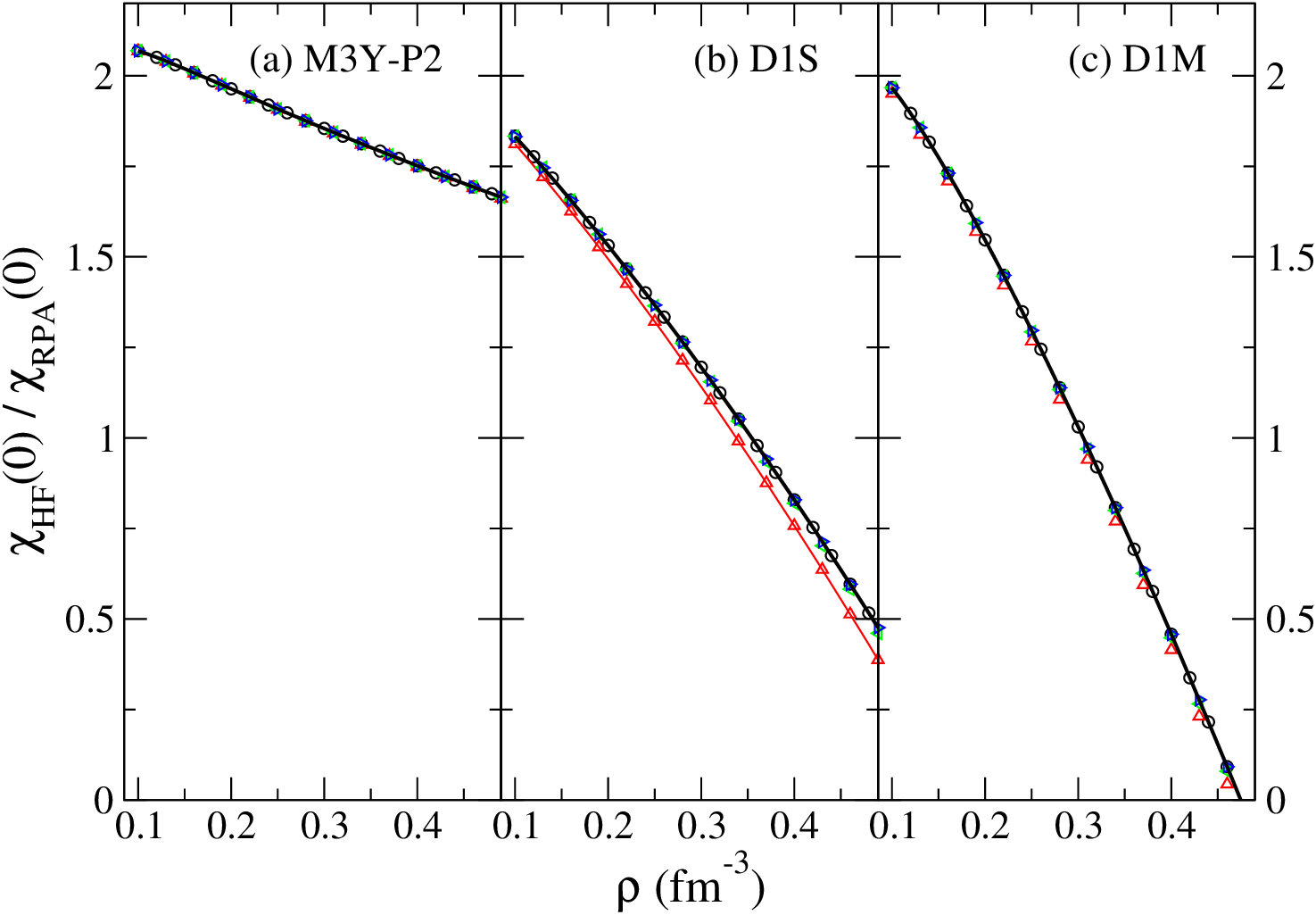}
\caption{Inverse of the spin susceptibility respect to the inverse Hartree-Fock spin susceptibility in the triplet isospin channel for symmetric nuclear matter as a function of density for the finite range effective
 forces including the tensor component: (a) M3Y-P2, (b) Gogny D1S and (c) Gogny D1M. Empty circles stand for the results taking into account only $G_0'$ in the Eq. \ref{suscept}, with all $H'$'s equal to zero. Then the different curves incorporate step by step the effects of the  $H'$'s. Curve with up triangles ($G_0', H_0'$), triangle oriented to the left ($G_0', H_0',H_1'$), triangles oriented to the right ($G_0',H_0',H_1',H_2'$) and the solid line stand for the full $RPA$ calculation.}
\label{FIG5}
\end{figure}

Results in the isovector channel in  symmetric nuclear matter are  shown  in Fig.~\ref{FIG5}
 for the same three finite-range effective interactions. The slope as a function of density is negative for all three effective forces, being larger for the Gogny forces. The effects of the
 tensor component are rather small and all the results obtained by successively turning on the 
different multipolar order of the Landau parameters stay very close to each other and to the 
calculations with only $G_0'$. In the case of D1M, the inverse of the spin susceptibility signals an instability that appears
at a density close to  $\rho = 0.45$ fm$^{-3}$. 

\begin{figure}[ht]
\centering\includegraphics[width=\linewidth]{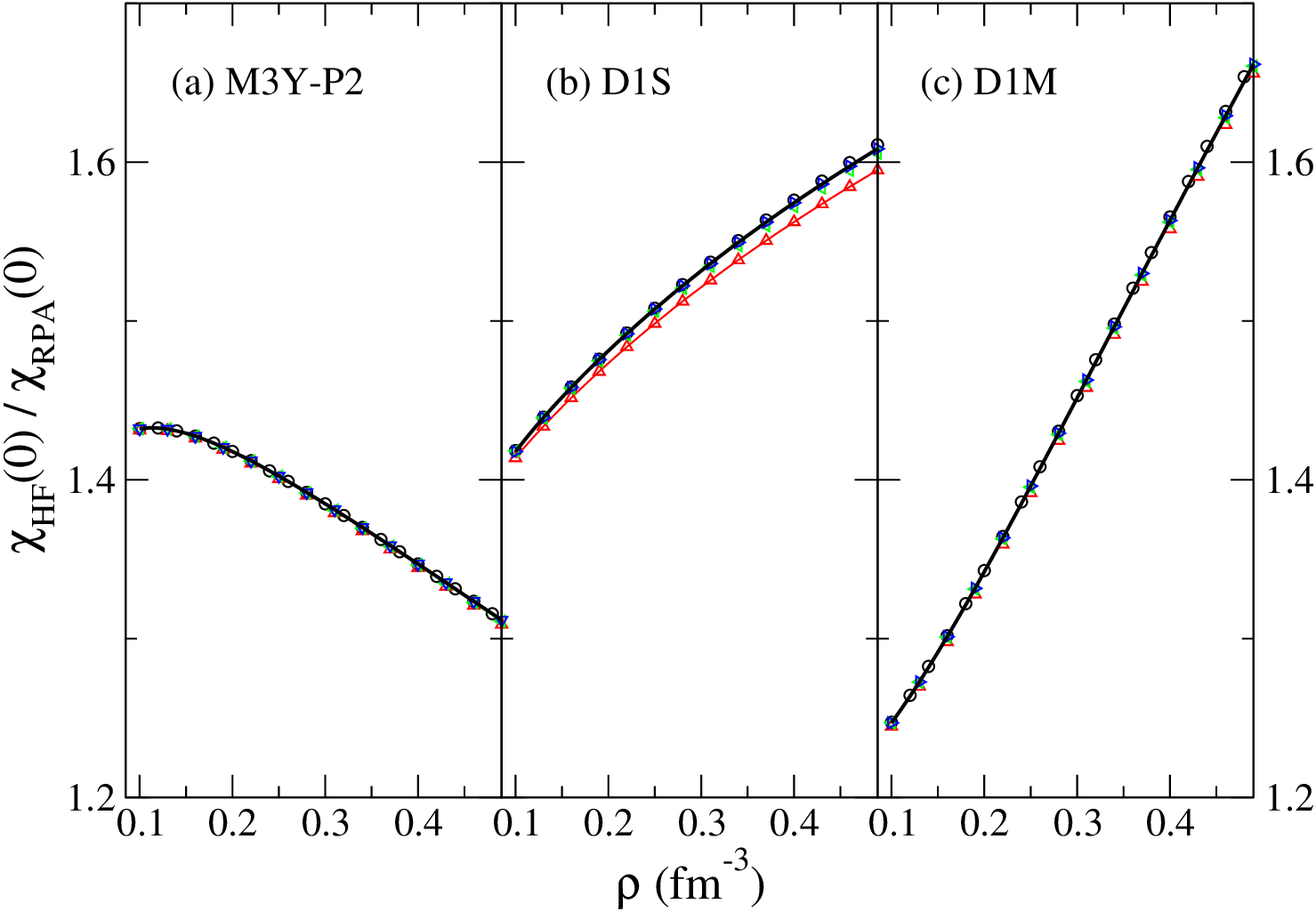}
\caption{Inverse of the spin susceptibility respect to the inverse Hartree-Fock spin susceptibility for neutron matter  as a function of density for the finite range effective forces including the tensor component:
(a) M3Y-P2, (b) Gogny D1S and (c) Gogny D1M. Empty circles stand for the results taking into account only $G_0^{(n)}$ in the Eq. \ref{suscept}, with all $H^{(n)}$'s equal to zero. Then the different curves incorporate step by step the effects of the  $H^{(n)}$'s. Curve with up triangles ($G_0^{(n)}, H_0^{(n)}$), triangle oriented to the left ($G_0^{(n)}, H_0^{(n)},H_1^{(n)}$), triangles oriented to the right ($G_0^{(n)},H_0^{(n)},H_1^{(n)},H_2^{(n)}$) and the solid line stand for
the full $RPA$ calculation.}
\label{FIG6}
\end{figure}

We turn now to the discussion of the results of the finite-range forces for neutron matter. The density dependence of the inverse of the spin susceptibility for neutron matter is shown in  Fig.~\ref{FIG6}.
 The effects of tensor force are  very small. The density slope is positive for the Gogny
 forces and no sign of instability is  present. In the case of M3Y-P2 the slope is negative
 but the value of the inverse magnetic  susceptibility remains always positive  with no sign
 of instability in the density region considered.   
For these forces we can conclude that the  combinations of $H_{\ell}^{(n)}$ parameters
 entering Eq.~(\ref{suscept}) are rather small and affect very little to the  behavior of
 the spin  susceptibility. In all cases, the results obtained by using only $G_0^{(n)}$ 
provide a  reliable evaluation of the spin susceptibility.

To understand better the size of the effects of the tensor Landau parameters in the
 spin susceptibility and to gain physical insight in the preceding discussions, in 
Table \ref{table1}  we report the neutron matter Landau parameters at $k_F=1.7$ fm$^{-1}$ for the different effective forces discussed along  the paper.

 As previously mentioned, concerning the Skyrme forces the only 
$H_{\ell}$ different from zero is  $\ell=0$. The  values of $H_0^{(n)}$ for T44 and T61
explain why the tensor effects are so small for T44 and sizeable for T61
(see Fig. \ref{FIG2}). In the case of  SLy5 and SIII,  both $H_0^{(n)}$ are 
negative and relatively large (see Fig. \ref{FIG1}), and the tensor effects are important. 

The finite range forces provide a relatively small values of 
$H_{\ell}^{(n)}$. This is particularly true for M3Y-P2, where the final result of the spin susceptibility is dominated 
by $G_0^{(n)}$. For the Gogny forces, although they are small they decrease very little with 
$l$ and as mentioned above, one should use all $H_{\ell}$ up to $l=3$ (which are the ones 
entering Eq. (\ref{suscept})) to have a converged result for the spin susceptibility.

\begin{figure}[ht]
\centering\includegraphics[width=\linewidth]{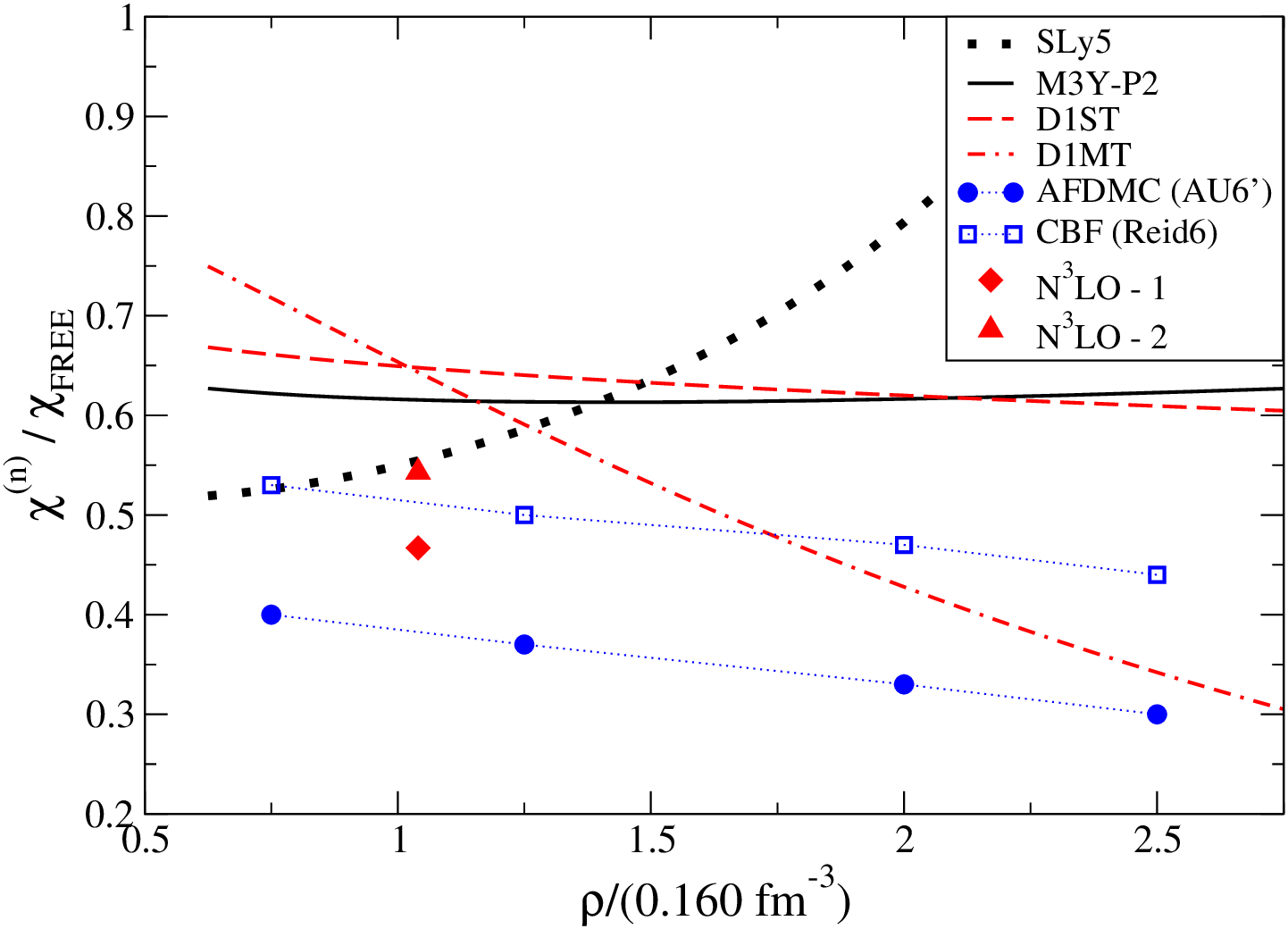}
\caption{(Color Online) Spin susceptibility of neutron matter measured respect to the free spin susceptibility 
 as a function of density for several interactions containing tensor forces.  SLy5 (squares),
M3Y-P2 (solid line), Gogny D1ST (dashed), Gogny D1MT (Dashed-dashed-dot), 
Av6'-AFDMC (full circles joint by dots) and Reid6-CBF (empty squares joint by dots).}
\label{FIG7}
\end{figure}

Finally, in Fig.~\ref{FIG7}  we show the spin  susceptibility of neutron matter in units
 of the free susceptibility as a function of density, for some of the effective 
interactions discussed in the paper and compare the results with microscopic calculations obtained with the Auxiliary Field Monte Carlo (AFMC) methods for the Argonne Av6'~\cite{fan01} and   results obtained within the framework of  Correlated Basis Functions (CBF) with the Reid-6 interaction \cite{kro92}.

The density dependence of these microscopic results
is smooth and similar for both cases, being the Monte Carlo results the lowest ones.
 Around  saturation density, both  effective  and realistic forces provide results relatively 
close. However the density dependence of the contact interaction (SLy5)  is completely different
 and grows very fast, in agreement with the instability shown  in Fig. \ref{FIG1}. 
The finite range of the effective interactions allows for a   more realistic density 
dependence, which is rather flat for M3Y-P2 and D1S-T and with a negative slope for D1M-T.

We also show  results at $k_F= 1.7$ fm$^{-1}$ obtained from the recently evaluated Landau parameters based on a microscopic calculation within the framework of chiral effective-field-theory~\cite{holt2013}. These authors considered the contributions from two-nucleon forces, derived at the $N^3LO$ approximation, and computed in a many-body perturbation  theory to  first
and second order. In addition they included also the leading-order from the $N^2LO$ chiral
 three-nucleon force. In any case, the  ph interaction at  the Fermi surface can be  expressed
 in terms of the Landau parameters.  Therefore, knowing 
the Landau parameters one can always use Eqs. (\ref{suscept}) to calculate the spin susceptibility. We calculate it using the Landau parameters obtained at first order (Table IV of this reference) from the $N^3LO$ interaction which is however submitted to a renormalization process in order to be used at this low-order in the perturbation theory. We also include the result using the Landau parameters reported in Table VI of this reference, which are calculated up to second-order in the perturbation
theory and include the effects of  $N^2LO$ chiral three-body forces. 
In both cases, the effects of the tensor on the spin susceptibility are very small, and the
results are safely reproduced using only $G_0^{(n)}$.  Around this density, all results 
are relatively  close. The first order calculation with $N^3LO$ produces a spin susceptibility
close to the Monte Carlo results with the Argonne potential. Taking into account the three-body forces and higher orders in the perturbation theory one gets a slighlty larger 
value of the spin-susceptibility closer to the finite range effective interactions.

\section{Summary and Conclusions}

We have critically analyzed the stability conditions of homogeneous nuclear and neutron matter against  several  low energy excitation  modes  of the Fermi sphere which are expressed in terms of the Landau parameters.  In particular, we have concentrated on the spin instabilities associated to spin 1. We have  explicitly calculated the  limit of zero-energy and zero-momentum transferred  of the spin response of nuclear matter and neutron matter  within  the Random Phase 
Approximation. This limit gives the spin  susceptibility of the system, which  can be expressed  in terms of the corresponding Landau parameters. One interesting conclusion, that to our knowledge has not been discussed before,  is  that the instability condition obtained  from the spin 
 susceptibility  is equivalent, {\it i.e.} provides the same critical density, to the one obtained 
 by the stability conditions emerging from the requirement  that the  ground state should 
 be stable respect to ph fluctuations with $J=1^+$. In the presence of the tensor force that corresponds to a coupling between fluctuations with quantum numbers $l=0$ and $l=2$. The effect of the contact tensor component  added to  Skyrme type effective forces is rather large  in the triplet spin channel in nuclear symmetric matter. It has been shown that   the  critical densities related to the spin instability  transition are smaller when the tensor component is taken into account.
 
In general, the finite range effective forces provide a spin susceptibility with less or no sign of  instabilities. At variance with the results obtained with Skyrme forces, all finite-range effective forces  considered in the paper provide Landau parameters such that the effect of the tensor component  in the spin susceptibility is very small. In the particular case of neutron matter, the density behavior of the spin susceptibility is close to the one emerging from the realistic forces. Finally, the results obtained from the Landau parameters provided by the Chiral forces for 
neutron matter at a density near  to the saturation density of nuclear matter are close to all
previous results around that density. The effects of the tensor force in the magnetic susceptibility when using these interactions are almost negligible.

\subsection*{Acknowledgments}
The discussions with D. Davesne and A. Pastore are gratefully acknowledged.
This work has been supported by MICIN (Spain), grants FIS2011-28617-C02-02 and  
FIS2011-24154 and Grant No. 2009SGR-1289 from Generalitat de Catalunya. Support 
of the Consolider Ingenio 2010 Programme CPAN CSD2007-00042 is also acknowledged.

\appendix

\section{}
As a previous step we require the knowledge of the functions:
\be
\alpha_i = \langle \cos^i\theta \ G_{HF} \rangle \, .
\ee
The limit $\omega=0, q=0$ of these integrals is easily calculated, with the result $\alpha_{2i+1} = 0$ and 
\be
\alpha_{2i} = - \,\frac{1}{2i+1} \, \frac{ N_0}{n_d}  \, .
\ee

Let us consider now the integrals over the momentum ${\bf k}_1$ entering the definition of 
 $K_i(n,M)$. They factorize in  an average over ${\bf k_1}$ of the propagator $G_{HF}$ multiplied by a certain number of spherical  harmonics $Y_{1,m}(\hat{k}_1)$ times an average over ${\bf k_2}$ of the propagator $G_{RPA}^{(M')}$ multiplied also by a certain number of spherical harmonics $Y_{1,m}(\hat{k}_2)$. First we pay attention to the average over ${\bf k_1}$. In this limit,  $\omega=0, q=0$, any integral containing an odd number of spherical harmonics, $Y_{1,m}(\hat{k}_1)$ , 
  reduces to a combination of $\alpha_{2i+1}$ and thus it is equal to zero. 
One has to consider only integrals containing an even number of spherical harmonics. As the propagator $G_{HF}$ is independent of the azimuthal angle $\phi$, the integrand reduces to the product of a certain number of factors $|Y_{1,1}|^2$ and $|Y_{1,0}|^2$. This fixes the parity of the index $n$ entering the definition of $K_i(n,M)$.

Once counted the number of times each of these terms appears, one has to write the integrand as products of $\sin^2 \theta$ and $\cos^2 \theta$ contained in the spherical harmonics. The integral over ${\bf k}_1$ reduces to a linear combination of  $\alpha_{2i}$, which  eventually is
 proportional to $\alpha_0$. Afterwards, the remaining integrals over ${\bf k}_2$ can also be grouped and simplified by taking adavantage of the relations $\sin^2+\cos^2=1$ or
$\sum_m |Y_{1m}|^2 = 3/4 \pi$. They reduce to either $\chi_{RPA}(0)$ or $S^{(M)}(0)$, {\it i.e.} the quantities we are interested in.  The final result is:
\bea
K_1(2n,M) &=& \frac{1}{2 n +1} \alpha_0 \chi_{RPA}^{(M)}(0) \, , \\
K_2(2n,M) &=&  \frac{1}{(2n+1)(2n+3)}  \alpha_0 \chi_{RPA}^{(M)}(0) + \frac{2n}{(2n+1)(2n+3)}  \alpha_0 S^{(M)}(0) \, , \\
K_3(2n+1,M) &=&   \frac{1}{2n+3}\alpha_0 S^{(M)}(0) \, , \\
K_4(2n+1,M) &=&   \frac{1}{2n+3}\alpha_0 S^{(M)}(0) \, , \\
K_5(2n,M) &=& \frac{1}{2 n +1} \alpha_0  S^{(M)}(0)  \, .
\eea

The Eqs.~(\ref{lachi}) and (\ref{laese}) reduce then to the closed system:
\bea
\chi_{RPA}^{(M)}(0) &=& \chi_{HF}(0) +
 \sum_n \left\{ \frac{A_{2n}}{2n+1} + \frac{B_{2n}}{(2n+1)(2n+3)} 
 \right\} \alpha_0  
\chi_{RPA}^{(M)}(0)  \nonumber \\
 &+&
\sum_n \left\{ \frac{2 n \, B_{2n}}{(2n+1)(2n+3)} - \frac{2 B_{2n+1}}{2n+3} 
+\frac{B_{2n}}{2n+1} \right\}  \alpha_0 S^{(M)}(0) 
\eea
\bea
S^{(M)}(0) &=& \frac{1}{3} \chi_{HF}(0) +
 \sum_n \left\{ \frac{A_{2n}+B_{2n}}{(2n+1)(2n+3)} -\frac{B_{2n+1}}{(2n+3)(2n+5)}  \right\} \alpha_0  
\chi_{RPA}^{(M)}(0) \nonumber \\
 &+&
\sum_n \left\{ \frac{2 n (A_{2n}+ B_{2n})}{(2n+1)(2n+3)} + \frac{B_{2n}- B_{2n+1}}{2n+3} 
- \frac{2 (n+1) B_{2n+1}}{(2n+3)(2n+5)} 
 \right\}  \alpha_0 S^{(M)}(0) \, .
\eea
The sums over $n$ have been done using MATHEMATICA, and one gets the results
 \bea
&&  \sum_n \left\{ \frac{A_{2n}}{2n+1} + \frac{B_{2n}}{(2n+1)(2n+3)} 
 \right\} = g_0-h_0+\frac{2}{3} h_1-\frac{1}{5} h_2 \nonumber \\
 &&\sum_n \left\{ \frac{2 n \, B_{2n}}{(2n+1)(2n+3)} - \frac{2 B_{2n+1}}{2n+3} 
+\frac{B_{2n}}{2n+1} \right\} = 3 h_0-2h_1+ \frac{3}{5} h_2 \nonumber \\
&&\sum_n \left\{ \frac{A_{2n}+B_{2n}}{(2n+1)(2n+3)} -\frac{B_{2n+1}}{(2n+3)(2n+5)}  \right\} 
= \frac{1}{3} g_0-\frac{1}{15} g_2 + \frac{1}{3} h_0-\frac{1}{15} h_1-\frac{1}{15} h_2 + \frac{1}{35} h_3 \nonumber \\
&&\sum_n \left\{ \frac{2 n (A_{2n}+ B_{2n})}{(2n+1)(2n+3)} + \frac{B_{2n}- B_{2n+1}}{2n+3} 
- \frac{2 (n+1) B_{2n+1}}{(2n+3)(2n+5)} \right\}
= \frac{1}{5} g_2+h_0-\frac{17}{15} h_1+\frac{3}{5} h_2-\frac{3}{35} h_3 \nonumber
 \eea
Note that in this limit, the contributions of Landau parameters with $\ell > 3$ cancel out exactly.

\vfill
\eject
\begin{table}
\begin{tabular}{ccrrrr}
\hline
                 &        &  0 & 1 & 2 & 3 \\
\hline
SIII             & F &   -0.864&  -0.297 && \\
                  & G &   -0.164 &  1.325 && \\
                  & H &  -0.779 &&& \\
\hline
SLy5          & F &  -0.251 & -1.173  && \\
                  & G &    0.099 &  1.173 && \\
                   & H &   -0.504 &&& \\
\hline
T44            & F &   -0.220 & -1.151 && \\
                 & G &   -0.373  & 1.643 && \\
                  & H &    0.027 &&& \\
\hline
T61            &  F &  -0.204 & -1.125 && \\
                  & G &   -0.839 &  2.104 && \\
                   & H &   -0.601 &&& \\
\hline
M3Y-P2             & F &   -0.557 & -0.366 & -0.176 & -0.108 \\
                 & G &   0.426  & 0.087  & 0.018 & -0.001 \\
                 & H &    0.026 &  0.049  & -0.498 &  0.034 \\
\hline
D1ST        & F &   -0.469 & -0.159  & -0.131 & -0.056 \\
                 & G &   0.462  & 0.171  & 0.129  & 0.054 \\
                 & H &   0.112 &  0.240 &  0.244 &  0.204 \\
\hline
D1MT       & F &   -0.354 & -0.474 &  0.187 &  0.082 \\
                 & G &    0.308 &  0.040  & 0.151  & 0.053 \\
                 & H &   0.090 &  0.200 &  0.219  & 0.194 \\
%\hline
%
%Chiral (IV) & F & -0.448  & -0.656 & -0.147 & -0.072 \\
%                 & G &  0.673  &  0.392 &  0.216 & 0.126 \\
%                 & H &  0.173  &  0.038 & -0.026  & -0.051\\
%\hline
%Low-k (IV) & F & -0.818  & -0.468 & -0.205 & -0.076 \\
%                 & G &  0.835  & 0.451  & 0.238  & 0.134 \\
%                 & H &  0.165  & 0.070  & -0.035 & -0.054 \\
\hline
\end{tabular}
\caption{Landau parameters for  neutron matter at  $k_F$=1.7 fm$^{-1}$ for the different
effective forces considered in the paper.}
\label{table1}
\end{table}

\newpage

\newpage

\newpage

\newpage

\newpage

\newpage

\end{document}